\documentclass{nature}
\usepackage{graphicx}
\usepackage[mathlines]{lineno}
\usepackage{amsmath}
\usepackage{url}
\usepackage{breakurl}

\title{Universal Scaling Laws in Metro Area Election Results} 

\author{Eszter Bok\'anyi$^{1\ast}$, Zolt\'an Sz\'all\'asi$^2$ \& G\'abor Vattay$^{1}$\\
\normalsize{$^{1}$Department of Physics of Complex Systems, E{\" o}tv{\" o}s University, Budapest, Hungary}\\
\normalsize{$^{2}$Children's Hospital, Harvard Medical School, Boston, Massachusetts, USA}\\
\normalsize{$^\ast$\emph{bokanyi@complex.elte.hu}}
}
\date{ }

\begin{document} 

\maketitle 


\begin{abstract}
We explain the anomaly of election results between large cities and rural areas in terms of urban scaling in the 1948-2016 US elections and in the 2016 EU referendum of the UK. The scaling curves are all universal and depend on a single parameter only, and  one of the parties always shows superlinear scaling and drives the process, while the sublinear exponent of the other party is merely the consequence of probability conservation. Based on the recently developed model of urban scaling, we give a microscopic model of voter behavior in which we replace diversity characterizing humans in creative aspects with social diversity and tolerance. The model can also predict new political developments such as the fragmentation of the left and 'the immigration paradox'.
\end{abstract}

Formation of cities is the result of socio-economic advantages of concentrating human populations in space outpacing associated costs.
A variety of disciplines including economics \cite{Henderson,fujita2001spatial,glaeser2011triumph}, 
geography \cite{batty2008size, isard1972location}, engineering\cite{kennedy2011evolution} and complex systems \cite{bettencourt2007growth,bettencourt2013origins,schlapfer2014scaling}
explain the existence of agglomeration or scaling effects and relate macroscopic properties of a city to its scale (population size).
Such relations are known across the sciences as scaling relations \cite{mitzenmacher2004brief}, and the systematic study of such relationships in cities is known as urban scaling. 
This scale-free, fractal-like behavior has also been observed in many human social networks \cite{newman2011structure} including cities \cite{batty2007cities,arbesman2011scaling,pan2013urban,sim2015great}. Using the population $N$ as the measure of city size, power law scaling takes the form 
\begin{linenomath*}
\begin{equation}
 Y=  Y_0\cdot N^\beta,   
\end{equation}
\end{linenomath*}
where $Y$ can denote material resources such as energy or infrastructure or measures of social activity such as wealth, patents and
pollution; $Y_0$ is a normalization constant. The exponent $\beta$ reflects general dynamic rules at play across the urban system.
Most urban socioeconomic indicators have superlinear $\beta>1$ exponents and as a result, larger cities are disproportionally the centers of innovation and wealth. Sublinear scaling $\beta < 1$ characterizes material quantities displaying economies of scale associated with infrastructure, where the agglomeration into cities pays off in having to provide fewer roads, shorter cables etc. Thus, material costs related to living in larger cities is disproportionally low.

Gomez-Lievano, Patterson-Lomba and Hausmann in Ref.\cite{hausmann} recently proposed a new model (GLPLH model) of superlinear scaling and demonstrated its validity on 43 urban phenomena related to employment, innovation, crime, education and diseases. The model accounts for the difference in scaling exponents and average prevalence across phenomena as well as for the difference in the variance within phenomena across cities of similar size.  The central idea is that a number $M$ of necessary complementary factors must simultaneously be present  for an urban phenomenon to occur.
The fraction of factors that an individual does not have and is expected to require from the city in order to be counted into a phenomenon is $q\in (0,1)$, and it quantifies the complexity of that phenomenon. The fraction of factors that a city provides for an individual is $r \in (0,1)$. It represents a measure of urban diversity and tends to accumulate logarithmically $r=a+b\cdot\log N$ with the population size, where $a$ and $b$ has been found to be constant across a wide range of urban phenomena. Alternatively, the fraction of factors {\em not} present in a city is $1-r=b\cdot\log N_0/N$,
where $\log N_0 =(1-a)/b$, and $N_0\approx  1.8\cdot 10^{14}$ is a hypothetical maximal diversity attainable in a city. Given a city with $m$ factors present, the probability that an individual requires any number of the $m$ factors that the city has, but none of the $M-m$ factors that the city does not have
is $P=(1-q)^{M-m}\approx e^{q(M-m)}$ for $q \ll 1$, and the average number of occurrence of the phenomena is $ Y=N\langle P\rangle_N,$ yielding
$Y \approx Ne^{qM(1-r(N))}=Ne^{qMb\log N_0/N}$
where we used $\langle e^{-qm} \rangle_N\approx e^{-Mr(N)}$ and 
averaging goes for cities of population $N$. Introducing the 
the scaling exponent $\beta=1+Mbq$, this scaling curve then takes the universal form
\begin{linenomath*}
\begin{equation}
 Y =N_0\left(\frac{N}{N_0}\right)^{\beta},\label{haus}
\end{equation}
\end{linenomath*}
where $N$ is now the part of population conceivably susceptible to the given urban phenomena. 

Scaling laws and universality have been observed in various aspects of the political process and elections \cite{bernardes2002election,mantovani2013engagement,PhysRevE.60.1067,chatterjee2013universality}, but the specific question of urban scaling of election results has not been addressed before. In the recent presidential elections in the US it has been noticed that votes for Democrats were disproportionally high in large cities \cite{florida}, and in the UK major cities also voted to remain in the EU. Here we show that election data in the US and in the UK show strong evidence of urban scaling.
Using the concept that tolerance and diversity are strongly coupled in cities\cite{florida2014rise}, we develop a microscopic model of voter behavior that produces the macroscopic level urban scaling, explains the observed single parameter scaling, and describes the distribution of deviations from the macroscopic curve.

\begin{figure}
\includegraphics[scale=1.2]{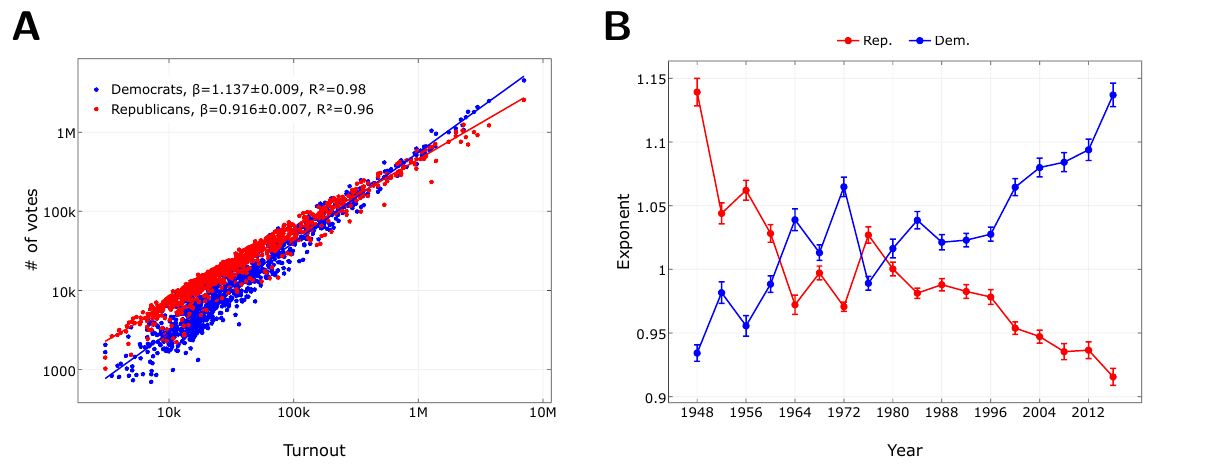} 
\caption{\textbf{Urban scaling in the presidential elections in the US.} {\bf A} Doubly logarithmic plot of votes cast for Republicans (red) and Democrats (blue) as
the function of the voter turnout for the  912 largest Metropolitan and Micropolitan Statistical Areas in 2016. Best fit line slopes $\beta$ and regression coefficients $R^2$ are in the insets. {\bf B} Scaling exponents for the Republicans (red) and Democrats (blue) with error-bars for the 18 presidential elections from 1948 to 2016.}
\end{figure}
 
First, we analyze data for the votes cast for the two main political parties in all post-World War II US presidential elections\cite{liep} and in the UK EU referendum \cite{ukelection}. In Fig. 1A we show votes for the political options as a function of voter turnout for the 912 largest Metropolitan and Micropolitan Statistical Areas representing about $82\%$ of the total voter population for the 2016 presidential election in the US (see supplementary material and methods for the UK). The votes for Democrats and 'Remain in the EU' scale superlinearly with exponents $\beta_{D}\approx 1.14$ and $\beta_{rem}\approx 1.09$, while votes for Republicans and 'Leave the EU' follow sublinear scaling with $\beta_{R}\approx 0.92$ and $\beta_{lea}\approx 0.91$, with high regression coefficients $R^2\ge 0.9$ indicating robust urban scaling. While the elections took place in two different political situations, nevertheless they show very similar exponents. 
In Fig. 1B we show the historical record of scaling exponents of the Democrats $\beta_{D}$ and of the Republicans $\beta_{R}$ for the 18 presidential elections in the period 1948-2016. The exponent of the Democrats has an increasing, while the exponent of the Republicans a decreasing historical trend. The Democrat and Republican curves roughly mirror each other in the whole period. The relation of the two exponents becomes apparent when we plot the Republican exponent as a function of the Democrat exponent in Fig. 2A. 

\begin{figure}
\includegraphics[scale=1.2]{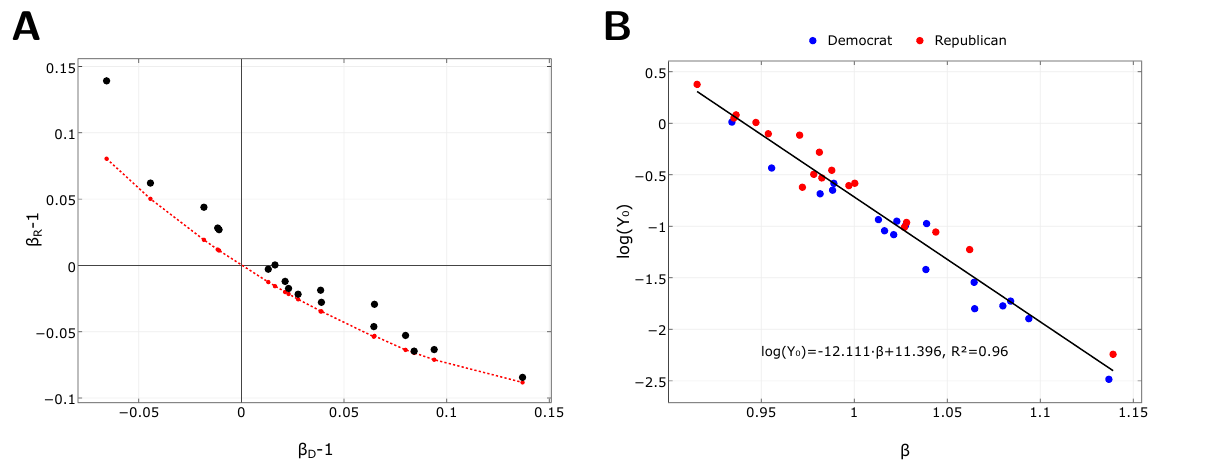}
\caption{\textbf{Interrelation of the parameters of urban scaling in US elections.}
{\bf A} Urban scaling exponents of Republicans as a function of the Democrats for 18 US presidential elections from 1948 to 2016 (dots) and the theoretical curve (red line) derived from probability conservation (\ref{betas}).
{\bf B} Intercepts of the scaling relations $\log Y_0$ as a function of the scaling exponent $\beta$ for Republicans (red) and Democrats (blue) for presidential elections in the period 1948-2016. Fitted line (\ref{line}) with parameters and regression coefficient in the inset.
}
\end{figure}

For each election and for each party we can determine the scaling exponent $\beta$ and the constant $Y_0$ independently from the fits. In Fig. 2B we plot $\log Y_0$ as a function of $\beta$. We find a very strong ($R^2=0.96$) linear relation
\begin{linenomath*}
\begin{equation}
 \log Y_0=-\alpha \beta +\delta,\label{line}
\end{equation}
\end{linenomath*}
for both parties and for all elections, with $\alpha=12.111$ and $\delta=11.396$. This indicates that the form of the scaling relation is independent of the party and election and has the universal form
\begin{linenomath*}
\begin{equation}
Y=e^{\delta-\alpha} N^* \left(\frac{N}{N^*}\right)^\beta\approx\frac{1}{2} N^* \left(\frac{N}{N^*}\right)^{\beta},\label{main}    
\end{equation}
\end{linenomath*}
where $N$ is the voter turnout in a city, $\beta$ is the exponent of the party and $\log N^*=\alpha$. The numerical factor $e^{\delta-\alpha}$ is equal to $1/2$ within numerical error and the parameter $N^*\approx 182.000$ is the average turnout of a US city of total population $429.000$ in 2016.  
The remarkable property of this scaling relation is that on average at turnout $N=N^*$ the parties share the votes equally
($Y_D=Y_R=N^*/2$) independent of their exponents $\beta_D$ and $\beta_R$ or of the year of the election and unaffected by historic changes in population. For cities above turnout  $N^*$ the party with higher $\beta$ gets the majority of votes, while below this turnout the party with smaller $\beta$ succeeds on average.
The observed linear relationship (\ref{line}) and the single parameter form (\ref{main}) of the scaling curve is predicted by the GLPLH model, therefore, it is reasonable to assume that it can be adapted to the election process. Formally, we recover our scaling curve (\ref{main}) from this theory by identifying the susceptible population with half of the voter turnout $N/2$ and by setting $N_0=N^*/2$. There are two discrepancies between our scaling curve (\ref{main}) and that of the GLPLH model. The GLPLH model is applicable for superlinear $\beta>1$ ($Mq>0$) values only, while in case of elections both superlinear and sublinear exponents arise, and the numerical value of $N_0$,  $\approx 1.8\cdot 10^5$ is nine orders of magnitude smaller for elections.

The main difference of elections from other urban phenomena is that the scaling curves influence each other via the competition for votes. 
This competition is expressed mathematically by the probability conservation  $Y_D/N+Y_{R}/N=1$ for the sum of the fraction of votes the parties get. Using (\ref{main}) and averaging for all cities yields
\begin{linenomath*}
\begin{equation}
\frac{1}{2}\left\langle\left(N/N^*\right)^{\beta_{D}-1}\right\rangle+\frac{1}{2}\left\langle\left(N/N^*\right)^{\beta_{R}-1}\right\rangle=1.\label{betas}
\end{equation}\end{linenomath*}
This equation guarantees that one of the exponents will be superlinear while the other sublinear (see supplementary materials and methods). Its numerical solution is shown in Fig. 2A. 
Thus, a model and an exponent derived for the results of one of the parties will determine the results of the other party via probability conservation.
The strategy of one of the parties will result in a superlinear exponent, which can be explained by a adaptation
of the GLPLH model, while the result of the party with the sublinear exponent is just a consequence of the other party's strategy. 

 A Scale-Adjusted Metropolitan Indicator\cite{bettencourt2010urban} (SAMI)
is the logarithmic deviation of the value $Y_i$ from the average scaling
curve for a city with population $N_i$ 
\begin{linenomath*}\begin{equation}
\xi_i=\log Y_i-\log Y_0-\beta\log N_i.\label{sami}
\end{equation}\end{linenomath*}
The GLPLH model predicts that SAMIs for a given city size range
are normally distributed, and their variance can be expressed with the complexity parameter $q$ and the number of complementary factors $M$ as
$\sigma^2_{SAMI}=q^2Mb(\log N_0 -\langle \log N \rangle),$ where $\langle \log N \rangle$ is the mean of the logarithm of city sizes. It can also be expressed with the scaling exponent
\begin{linenomath*}\begin{equation}
\sigma^2_{SAMI}=q(\beta-1)(\log N_0 -\langle \log N \rangle).
\end{equation}\end{linenomath*}

\begin{figure}
\centerline{
\includegraphics[scale=1.2]{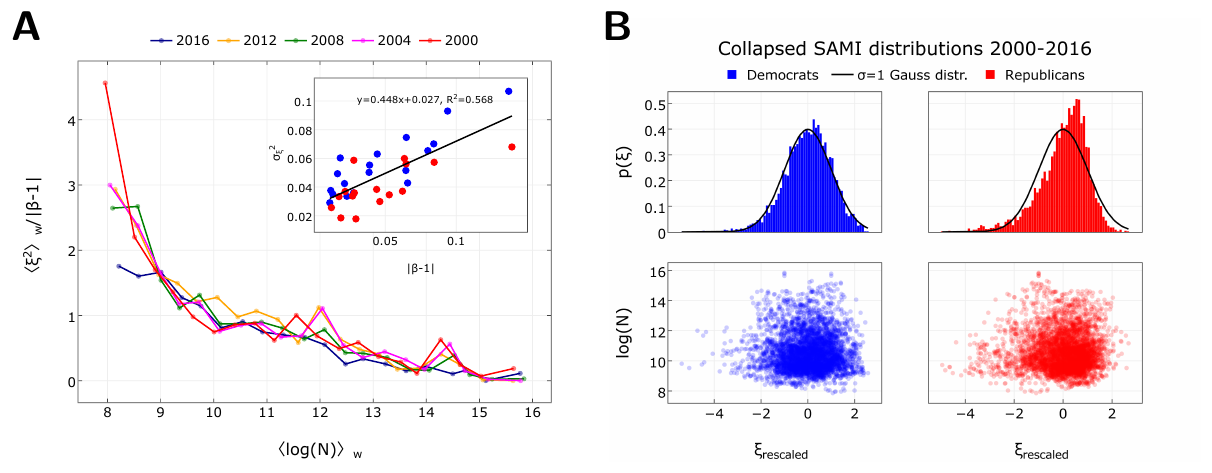}} 
\caption{\textbf{Fluctuations around the average scaling curve.}
{\bf A}, Variance of the deviation from the average scaling curve as a function of the logarithmic city size, measured in voter turnout. City sizes are binned into 20 windows of uniform sizes on logarithmic scale. In the inset, standard deviation of SAMIs (\ref{sami}) for all metropolitan areas in our study as a function $\beta-1$.
Best fit line parameters are in the inset.
{\bf B}, Standardized deviation of SAMIs for the last five US presidential elections. Lower panel: Scatter plot for the
Democrat (blue) and Republican (red) standardized deviations (horizontal axis) and logarithmic city size (vertical axis). Upper panel:
Distribution of the standardized deviations, x axis is shared with that of the the lower panel. For the Democrats (blue) it is nearly a standard normal distribution (solid line). For Republicans (red) it is a skewed distribution deviating from the standard normal distribution (solid line).
}
\end{figure}

In Fig. 3 we check the general validity of this formula for both parties and for all elections in the 1948-2016 period. For the variance averaged over all metropolitan areas we can confirm the proportionality with $\beta-1$ (see inset of Fig. 3A), indicating also that the complexity parameter $q$ is approximately constant. From the fitted line and from the numerical value $ \langle \log N \rangle=10.55$ for the 2016 election, we get $q\approx 0.28$. Then, in the 2000-2016 period for the superlinearly scaling results of the Democrats, we can make a more detailed calculation for ten windows of city sizes. In Fig. 3A (main) we can see that curves of $\sigma^2_{SAMI}/(\beta-1)$ for different elections in these windows collapse onto the same curve confirming that the complexity parameter is constant. This implies that the change of
the scaling exponent $\beta_D$ for the Democrats in this period comes solely from the change of the number of complementary factors $M$.
Deviations of cities from the average scaling curves can then be 
standardized in the windows using the window-wise variances. In Fig. 3B we show the distribution of these standardized SAMIs for both parties. As expected, the distribution of these standardized SAMIs for Democratic party is standard Gaussian, in agreement with the GLPLH model. However, the same procedure results in a skewed distribution for the Republicans. Their distribution is not normal and the GLPLH model doesn't apply which again confirms that the two parties don't have an equal role in the urban scaling.

Now, the question arises: in the context of elections, what are the necessary complementary factors that must simultaneously be present in order to vote for Democrats in the 1988-2016 period, where their exponent is superlinear? We found that the complexity parameter $q$ is approximately constant, so from the
4-6 times growth of $\beta_{D}-1$ in this period we can conclude that the number of factors $M$ got multiplicated about 4-6 times. The concrete value of $M$ cannot be determined from our data, only the product $bM$ that changed from about $0.09$ to $0.52$ with $b$ being constant. If we could find these factors, then Republicans (or the Leave campaign) could be characterized as comprising of
those voters, who don't accept at least one of those $M$ factors.
Here, we can just conjecture based on the agenda of Democrats, that these are "liberal values" in general, and the Democrat voter typically accepts all of these values simultaneously. Such values include tolerance and acceptance towards various social groups ranging from women and blacks at the beginning and middle of the $20^{\mathrm{th}}$ century to LGBT communities, immigrants, refugees and various other social minorities recently. In case someone is not able to accept at least one of these, then that person will probably not vote for the Democrats. That explains why groups of the Republican and the Leave voters look so heterogeneous: they consist of groups that oppose at least one of these liberal values, and that are not held together by a common political agenda otherwise. 
In this context, we can identify $q$ as a probability that a voter -- left on its own devices -- rejects one of the $M$ liberal values, and $r(N)$ is the probability that a city of size $N$ makes a voter tolerant towards those values. Social diversity grows with the city size and voters in cities can face an increasing number of social issues and can develop tolerance towards them. This is in accordance with  'the immigration paradox' in Britain, where voters living near immigrants develop a tolerance, while
those who do not are more likely to reject them\cite{goodwin20162016}.
Therefore, we expect that just like other types of diversities in cities,
tolerance grows like $r(N)\sim \log N/N_0$, but the number of maximal social diversity is reached at $N_0\approx 4\cdot 10^5$, which is smaller than the diversity $N^*\approx 1.8 \cdot 10^{14}$ observed for the more general type of diversity, characterizing humans in creative aspects. Finally, there is one more consequence of this model: as the number of liberal values $M$ seems to grow continuously, the potential voters who don't accept one of them also increases, and becomes detrimental for electoral success. This leads to the fragmentation of the political left, since a larger number of smaller parties accepting only a subset
of the $M$ values, or even "single-issue" parties can minimize the number of estranged voters and maximize the aggregated votes of all these parties. We believe that he model and the calculations could further be extended to metropolitan areas in other countries or to electoral systems with multiple choices.



\section*{Acknowledgements}
Authors thank the Hungarian National Research, Development and Innovation Office under Grant No. 125280 and the grant of Ericsson Inc.

\section*{Author contributions}

E.B. analyzed the data, G.V. conceived the study, E.B., G.V. and Z.Sz. wrote the manuscript.

\section*{Competing interests}

The authors declare no competing financial interests.

\clearpage

{\Large\bfseries\noindent\sloppy \textsf{Supplementary Materials and Methods} \par}


\section{Data sources}

We downloaded county-level historical US presidential election datasets from \cite{liep}. We calculated the total number of votes for the Democratic and Republican Party and the turnouts for all Metropolitan and Micropolitan Statistical Areas \cite{msa} by matching MSA's to the county level data \cite{cbsatofips}.

As for the UK, we downloaded electorate-level number of votes for Remain and Leave from the EU referendum result dataset \cite{uk}. We filtered the UK electorates based on whether they have a city in their core \cite{ukcities}, because the resolution of the data available about the referendum was not enough to consider using cities as units.

\section{Data fit}

For each year $y$, we assume that the expected value of the number of voters for a party ($D$, Democrat or $R$, Republican) scales with the size of a city in the following way:

\[Y^{(y)}(N)=Y_{0,D/R}^{(y)}\cdot N^{\beta^{(y)}_{D/R}}.\]

Taking the logarithm of both sides, we can fit a line using OLS fit on the $(\log Y, N)$ pairs for each election for both parties (we leave the year and party notations for simplicity reasons):

\[\log(Y(N))=\log(Y_0)+\beta \cdot \log(N),\]

where the $\beta$ denotes the slope, $\log Y_0$ the the intercept of the fitted line, thus $\beta$ is the exponent of the party in year $y$.

\section{Pivotal point}

If we assume that the intercept $\log(Y_0)$ is a function of $\beta$ that changes slowly with $\beta$, and we know that $\beta$ is always close to 1, then we can approximate $\log Y_0$ around 1 linearly:
\[\log(Y_0(\beta))\approx \underbrace{\log(Y_0(1))}_{\delta-\alpha}+(\beta-1)\underbrace{\left.\frac{\partial \log(Y_0(\beta))}{\partial\beta}\right|_{\beta=1}}_{-\alpha}+\dots=-\alpha\cdot\beta+\delta\]

In the case of $\beta=1$, it has to be true, that
\[Y_0(1)=e^{\delta-\alpha}=\langle p\rangle=p_0,\]
the city-averaged voter fractions, because that would mean that every city votes as if all voters were dispersed homogeneously:
\[Y_D(N)=p_0N.\]

Let $\alpha=\log N^*$, then $p_0=e^\delta/e^\alpha=e^\delta/N^*$.

\[\log(Y_0(\beta))=-\log N^*\cdot\beta+\log(p_0)+\log N^*\]

By substituting it into the original scaling relation:

\[\log Y (N) = \log (p_0N^*)-\beta\cdot\log N^*+\beta\cdot \log N\]
thus,
\[Y(N)=p_0N^*\left(\frac{N}{N^*}\right)^\beta=p_0N\left(\frac{N}{N^*}\right)^{\beta-1}\]

This implies that all fitted lines have to go through the ($N^*$, $p_0N^*$) point, because at $N=N^*$, $Y$ equals to $p_0N^*$ regardless of the value of $\beta$. Also note, that $N^*$ is universal for both parties and for all elections. Thus, the scaling relations only have only parameter, the scaling exponent $\beta$.

\section{Exponent relationship}

In a given year, for every city $i$ it holds that the number of Democrat and Republican voters is approximately equal to the turnout in the city:

\[\frac{Y^{(i)}_D}{N^{(i)}}+\frac{Y^{(i)}_R}{N^{(i)}}=1\]

Assuming scaling, the expected values of the Democrat and Republican voters can be substituted:

\begin{align*}
Y^{(i)}_D&=\frac{1}{2}N^*\left(\frac{N^{(i)}}{N^*}\right)^{\beta_D}=\frac{1}{2}N^{(i)}\left(\frac{N^{(i)}}{N^*}\right)^{\beta_D-1}\\
Y^{(i)}_R&=\frac{1}{2}N^*\left(\frac{N^{(i)}}{N^*}\right)^{\beta_R}=\frac{1}{2}N^{(i)}\left(\frac{N^{(i)}}{N^*}\right)^{\beta_R-1}
\end{align*}

Thus,

\begin{align*}
\frac{1}{2}\left(\frac{N^{(i)}}{N^*}\right)^{\beta_D-1}+\frac{1}{2}\left(\frac{N^{(i)}}{N^*}\right)^{\beta_R-1}&=1\\
\left(\frac{N^{(i)}}{N^*}\right)^{\beta_D-1}+\left(\frac{N^{(i)}}{N^*}\right)^{\beta_R-1}&=2
\end{align*}

Because the exponents $\beta_D$ and $\beta_R$ are close to 1, the left hand side can be approximated to the second order

\begin{align*}&1+
(\beta_D-1)\cdot\log\frac{N^{(i)}}{N^*}+
\frac{1}{2}(\beta_D-1)^2\cdot\left(\log\frac{N^{(i)}}{N^*}\right)^2+
\dots+\\
&1+
(\beta_R-1)\cdot\log \frac{N^{(i)}}{N^*}+
\frac{1}{2}(\beta_R-1)^2\cdot\left(\log \frac{N^{(i)}}{N^*}\right)^2+
\dots
=2
\end{align*}

\clearpage

Let us average the equation over all cities in a year:

\begin{align*}
&(\beta_D-1)\cdot\left<\log\frac{N^{(i)}}{N^*}\right>+
\frac{1}{2}(\beta_D-1)^2\cdot\left<\left(\log\frac{N^{(i)}}{N^*}\right)^2\right>+\\
&(\beta_R-1)\cdot\left<\log \frac{N^{(i)}}{N^*}\right>+
\frac{1}{2}(\beta_R-1)^2\cdot\left<\left(\log \frac{N^{(i)}}{N^*}\right)^2\right>
=0
\end{align*}

In the first order, $\beta_R-1=-(\beta_D-1)$. Because the term $(\beta_R-1)^2$ is small, we only use its first order approximation, thus:

\begin{align*}
&(\beta_D-1)\cdot\left<\log\frac{N^{(i)}}{N^*}\right>+
\frac{1}{2}(\beta_D-1)^2\cdot\left<\left(\log\frac{N^{(i)}}{N^*}\right)^2\right>+\\
&(\beta_R-1)\cdot\left<\log \frac{N^{(i)}}{N^*}\right>+
\frac{1}{2}(-(\beta_D-1))^2\cdot\left<\left(\log \frac{N^{(i)}}{N^*}\right)^2\right>
=0
\end{align*}

\[
(\beta_D-1)\cdot\left<\log\frac{N^{(i)}}{N^*}\right>+
(\beta_D-1)^2\cdot\left<\left(\log\frac{N^{(i)}}{N^*}\right)^2\right>+
(\beta_R-1)\cdot\left<\log \frac{N^{(i)}}{N^*}\right>
=0
\]

\[
\beta_R-1=
-(\beta_D-1)\cdot+
(\beta_D-1)^2\cdot\frac{\left<\left(\log\frac{N^{(i)}}{N^*}\right)^2\right>}{\left<\log\frac{N^{(i)}}{N^*}\right>}
\]

\section{EU referendum UK 2016}

Similarly to that of the presidential election dataset in the United States, we fitted the $Y=Y_0\cdot N^\beta$ function on the number of Remain and Leave votes for the EU referendum in the cities of the United Kingdom. Electorate-level data was obtained from the homepage of the Electoral Commission \cite{ukelection}. Since we did not have a city-level resolution, we took electorates that were centered around a city, and used only their turnouts as $N$, and number of voters as $Y$.

Because the distribution of city sizes in the UK is very uneven even on the logarithmic scale with London being disproportionally large, we weighted the points by $1/N$ in the OLS fit on the double logarithmic plot.

As in the case of the US Democrats, the Remain votes showed a strong superlinear scaling with $\beta_{\mbox{Remain}}=1.08$, while the Leave votes scale sublinearly $\beta_{\mbox{Leave}}=0.91$

\begin{figure}
    \centering
    \includegraphics[width=0.7\textwidth]{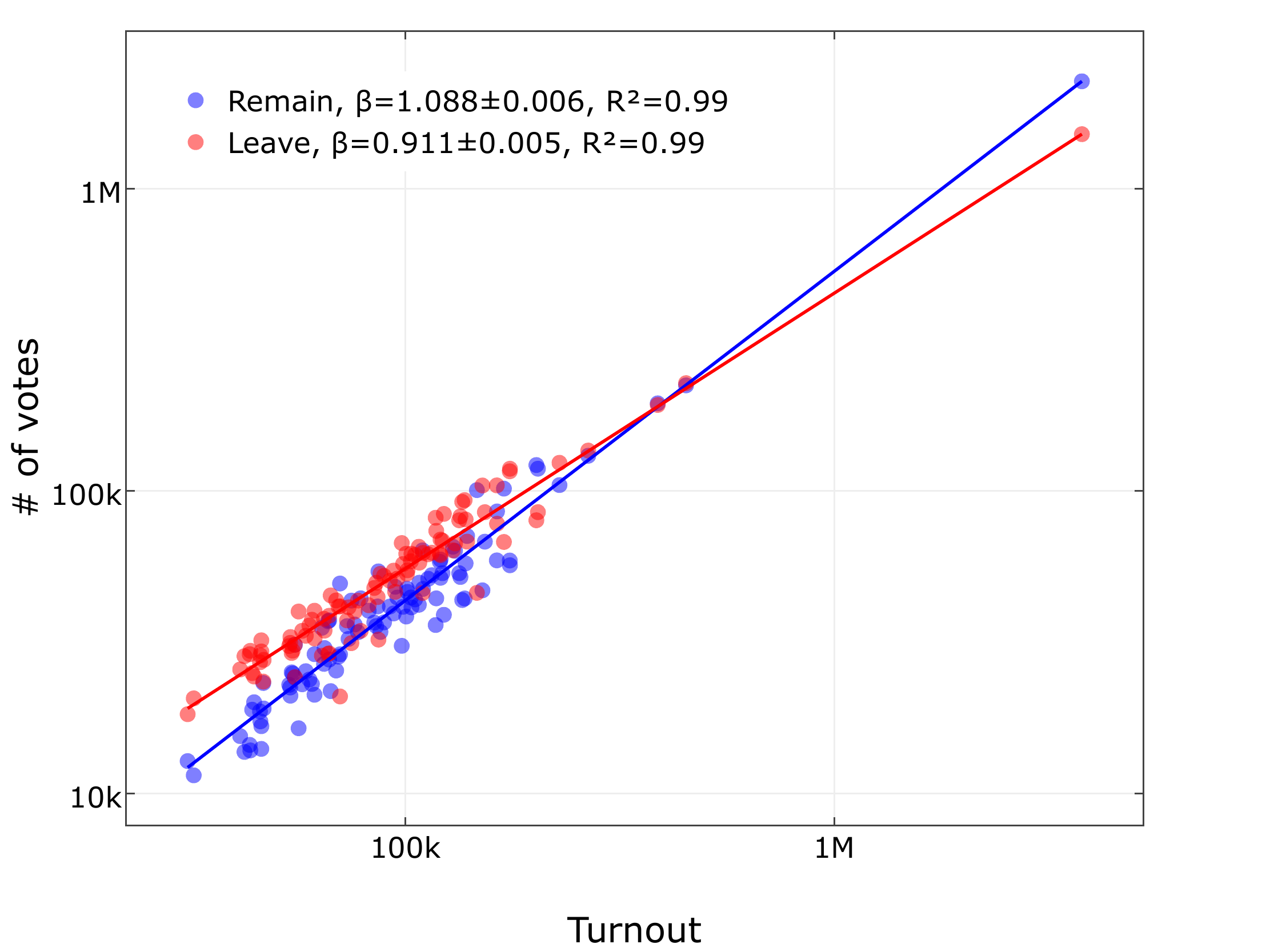}
    \caption{Urban scaling in the EU referendum results in the UK. Doubly logarithmic plot of votes cast for Leave (red) and Remain (blue) as the function of the voter turnout for the electorates surrounding cities. Best fit line slopes $\beta$ and regression coefficients $R^2$ are in the insets. }
    \label{fig:brexit}
\end{figure}

The results suggest that a similar mechanism can be behind this phenomenon, as behing the Democrat voters in the US.


\end{document}